\documentclass[a4paper,fleqn,usenatbib]{mnras}
\usepackage{amssymb,amsmath,graphicx,psfrag,mathtools,bm,amsbsy}
\usepackage[T1]{fontenc} 
\usepackage{ae,aecompl} 
\usepackage{url}
\title[FRB and the Cosmological Principle]{FRB Strength
Distribution Challenges the Cosmological Principle}
\author[J. I. Katz]{
J. I. Katz,$^{1}$\thanks{E-mail katz@wuphys.wustl.edu} 
\\
$^{1}$Department of Physics and McDonnell Center for the Space Sciences,
Washington University, St. Louis, Mo. 63130 USA 
}
\date{Accepted XXX.  Received YYY; in original form ZZZ} 
\pubyear{2017} 
\date{\today}
\begin{document} 
\label{firstpage} 
\pagerange{\pageref{firstpage}--\pageref{lastpage}} 
\maketitle 
\begin{abstract}
The distribution of FRB fluxes and fluences is characterized by a few very
bright events and a deficiency of fainter events, compared to expectations
for a homogeneous space-filling distribution.  I define a metric to quantify
this, and apply it to the 17 presently known Parkes FRB, products of a 
comparatively homogeneous search.  With 98\% confidence we reject the
hypothesis of a homogeneous distribution in Euclidean space.  Possible
explanations include a reduction of fainter events by cosmological redshifts
or evolution or a cosmologically local concentration of events.  The former
is opposed by the small value of the one known FRB redshift.  The latter
contradicts the Cosmological Principle, but may be explained if the brighter
FRB originate in the Local Supercluster.
\end{abstract}
\section{Introduction}
From the discovery of the first Fast Radio Burst \citep{L07}, it has been
noticed that there is a deficiency of weaker bursts compared to the number
$N \propto S^{-3/2}$ expected in a Euclidean Universe.  In order to have a
reasonably homogeneous statistical sample we consider only the 17 bursts
observed at Parkes out of the 23 bursts in the FRB Catalogue \citep{P16}.
The more recent discoveries of very bright bursts by UTMOST \citep{C17} and
ASKAP \citep{B17} were made with instruments of lower sensitivity and
cannot be commingled with Parkes observations in a homogeneous data set.

We wish to test the hypothesis that FRB are homogeneously distributed in a
Euclidean Universe.  This cannot be exactly correct because we know that the
Universe is not Euclidean and evolves.  However, we are ignorant of the
evolution of the FRB source population and of their spectra (required to
calculate K-corrections), so the Euclidean model is as good as any we could
choose, and has the advantage of specificity.  The one known FRB redshift is
small (0.193; \citet{T17}), suggesting that the Euclidean model is in fact a
fair approximation.  The statistics of the fainter Parkes FRB
\citep{K16a,K16b} are approximately consistent with the Euclidean model, but
this has not been demonstrated quantitatively.  This paper develops a
quantitative metric and applies it to the Parkes dataset.
\section{The Metric}
The assumption of sources homogeneously distributed in Euclidean space makes
definite predictions.  We consider the $\alpha$-th moment of the {
received signal} $S$, where $S$ may be any quantity that satisfies an
inverse square law and has a definite detection threshold $S_0$.  Examples
include flux, fluence and flux times the square root of the pulse width, as
for UTMOST \citep{C17}.  This last quantity is appropriate when the signal
must be distinguished from detector thermal noise.  The detection threshold
may depend non-trivially on the pulse shape and other quantities that (if
cosmological redshift is small) do not depend on distance.

The normalized $\alpha$-th moment of $S$ is defined
\begin{equation}
f(\alpha) \equiv \langle S^\alpha \rangle = {\int S^\alpha dN \over
\int dN},
\end{equation}
where $N$ is the number of sources in a catalogue.  { For a homogeneous
and continuous distribution of sources of number density $n({\cal L})$ per
unit source strength $\cal L$, $dN = 4 \pi n({\cal L}) R^2 dR d{\cal L}$,
where $R$ is their distance and $\cal L$ is luminosity for steady sources,
energy for temporally unresolved bursts and something more complicated but
still following the inverse square law (if cosmologically local, and
ignoring any effect of intergalactic dispersion on detectability) for
temporally resolved bursts.  Then $S = S_0 R_0^2/R^2$, where
$R_0(\Omega,{\cal L})$ is the distance at which a source of strength
$\cal L$ in the direction $\Omega$} is at the detection threshold.  {
$R_0(\Omega,{\cal L})$ depends on the unknown distance of the source from
the axis of the telescope beam.  Integrating,}
\begin{equation}
\label{Salpha}
\begin{split}
f(\alpha) &= {\int\!d\Omega \int\!S^\alpha dN \over \int\!d\Omega \int\!dN}\\
          &= {\int\!d\Omega\int\!d{\cal L}\int_0^{R_0(\Omega,{\cal L})}\!
             S_0^\alpha[R_0(\Omega,{\cal L})/R]^{2\alpha} n({\cal L}) R^2\,dR
             \over \int\!d\Omega\int\!d{\cal L}\int_0^{R_0(\Omega,{\cal L})}
             \!n({\cal L}) R^2\,dR}\\
          &= {3 \over 3 - 2\alpha} S_0^\alpha {\int\!d\Omega\int\!d{\cal L}
             \,n({\cal L})R_0^3(\Omega,{\cal L}) \over
             \int\!d\Omega\int\!d{\cal L}\,n({\cal L})R_0^3(\Omega,{\cal L})}
           = {3 \over 3 - 2\alpha} S_0^\alpha.
\end{split}
\end{equation}
{ The angular dependence of telescope sensitivity cancels, provided $S_0$
is, at least statistically, independent of $\Omega$.  This is expected
because the signals are processed and analyzed without knowledge of
$\Omega$.}  $f(\alpha)$ is meaningful only for $\alpha < 3/2$ and useful
only for $\alpha > 0$.

In general, $S_0$ is poorly known { because of its complex dependence on
pulse width and profile}, so that it is not possible to use Eq.~\ref{Salpha}
directly.  Instead, define a metric
\begin{equation}
\label{metric}
F(\alpha_1,\alpha_2) \equiv {f(\alpha_1) \over
f(\alpha_2)^{\alpha_1/\alpha_2}} = 3^{1-\alpha_1/\alpha_2}
{(3-2\alpha_2)^{\alpha_1/\alpha_2} \over 3-2\alpha_1}.
\end{equation}
This is dimensionless and independent of our knowledge or ignorance of
$S_0$.  A simple and intuitively appealing choice of parameters is $\alpha_1
= 1$ and $\alpha_2 = 1/2$, for which a homogeneous source distribution in
Euclidean space yields
\begin{equation}
F(1,1/2) = {4 \over 3}.
\end{equation}
\section{Finite Sample Statistics}
\begin{figure}
\centering
\includegraphics[width=0.99\columnwidth]{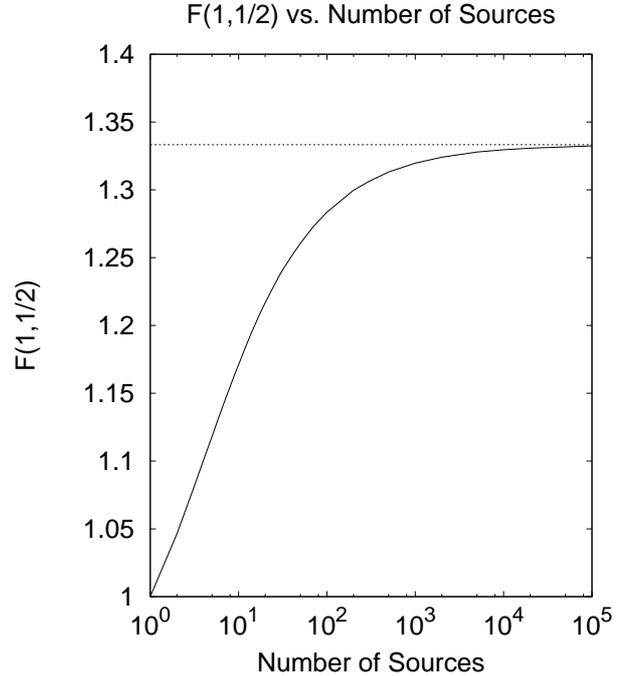}
\caption{\label{FFvN} Mean $F(1,1/2)$ as a function of the number $N$ of
sources.  For $N \to \infty$, $F(1,1/2)$ slowly converges to the continuum
limit of 4/3 shown by the dotted line.}
\end{figure}

The preceding results apply to continuously distributed sources.  In
practice, sources are discrete and catalogues are finite, so the predicted
values of $F$ and their distribution must be calculated by Monte Carlo
methods.  $N$ sources are randomly but statistically uniformly distributed
within a sphere whose outer radius is their detection limit.
The mean value of $F$ as a function of $N$ is shown in Fig.~\ref{FFvN} based
on $10^6$ realizations.  It approaches 4/3 only slowly as $N \to \infty$
because the inverse square law gives the poorly sampled small fraction of
sources close to the observer a disproportionate influence.

\begin{figure}
\centering
\includegraphics[width=0.99\columnwidth]{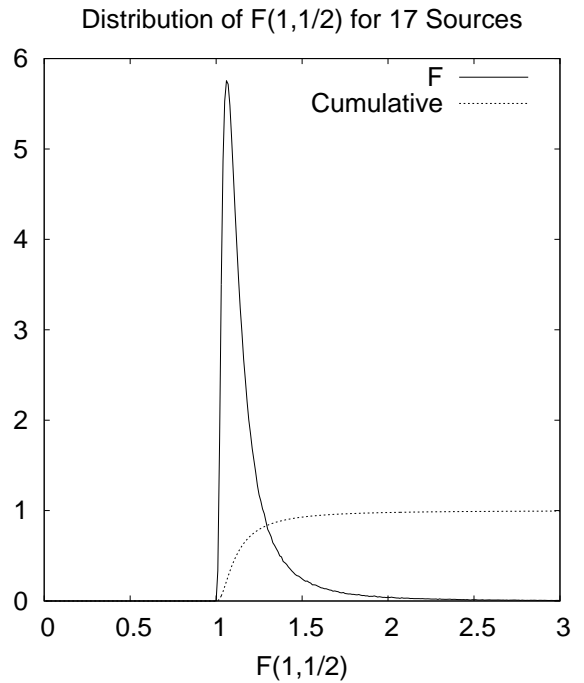}
\caption{\label{FFdistrib} Normalized distribution of $F(1,1/2)$ for a 17
element catalogue.  The normalized cumulative distribution is also shown.}
\end{figure}

The distribution of $F(1,1/2)$ over $10^6$ Monte Carlo realizations of 17
sources, corresponding to the FRB catalogue used in Sec.~\ref{data}, is
shown in Fig.~\ref{FFdistrib}.  The distribution is narrowly peaked but very
skew, with a maximum at 1.06, a mean of 1.21 and a standard deviation of
0.12, although it is far from Gaussian.  As implied by Fig.~\ref{FFvN}, this
distribution is only weakly dependent on $N$.

\begin{figure}
\centering
\includegraphics[width=0.99\columnwidth]{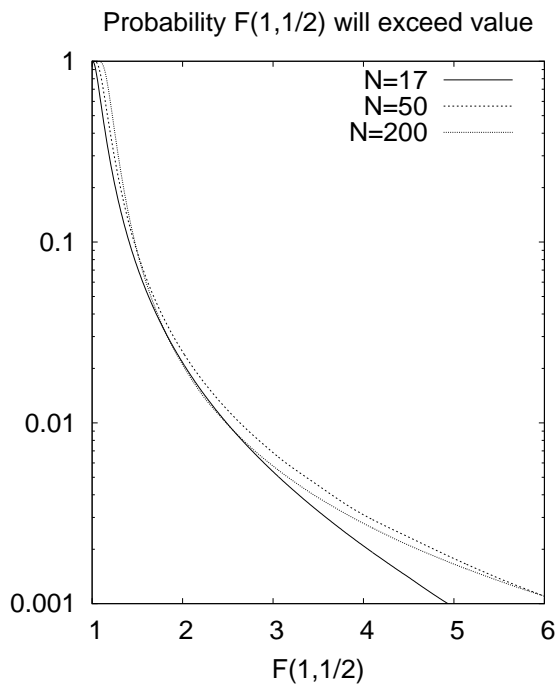}
\caption{\label{FFcum} The probability that for sources homogeneously
distributed in Euclidean space $F(1,1/2)$ will exceed the indicated value
for catalogues with 17, 50 and 200 entries, based on $10^8$ Monte Carlo
realizations.}
\end{figure}

For testing the significance of a value of $F$ larger than the predicted
mean the long tail of the cumulative distribution must be examined in more
detail.  This is shown in Fig.~\ref{FFcum}.
\section{Application to FRB}
\label{data}
We consider three possible definitions of the Parkes FRB detection
threshold: the flux, the fluence times the square root of the pulse width
$W$ (as expected when detection is only limited by thermal noise in the
detector) and the fluence.  The results are shown in Table~\ref{FRBcat}.
\begin{table}
\centering
\begin{tabular}{|c|ccc|}
\hline
Catalogue & Flux & Flux$\times W^{1/2}$ & Fluence \\
\hline
17 Bursts & 3.29 & 2.49 & 2.37 \\
{ 17 Randomized} & $3.31 \pm 0.26$ & $2.50 \pm 0.16$ & $2.38 \pm 0.22$ \\
{ 16 Bursts} & { 2.01} & { 2.23} & { 2.57} \\
15 Bursts & 1.12 & 1.06 & 1.10 \\
\hline
\end{tabular}
\caption{\label{FRBcat} $F(1,1/2)$ for the 17 Parkes FRB in the FRB
Catalogue \citep{P16}, { for the 16 FRB excluding the most recent and
very bright FRB 150807 and} for the 15 FRB remaining after removal of the
two brightest bursts FRB 010724 and FRB 150807, for three possible measures
of signal strength.  { Comparison of the 17 and 16 Bursts entries limits
the possible bias introduced by initiating this study after the discovery of
a second bright Parkes burst.  Randomized describes a synthetic catalogue in
which 15 weak bursts are chosen randomly from the 15 actual weak bursts;
some may be omitted and others represented multiple times, while the two
strong bursts are always included.  This simulates possible effects of
variations in detection thresholds and also estimates the small-sample
uncertainties ($1\sigma$) of the calculated $F(1,1/2)$.}}
\end{table}

Comparing to Fig.~\ref{FFcum} and more detailed tabular data, the Euclidean
hypothesis may be rejected at the 98\% level for a fluence threshold and
at even higher levels of confidence for other assumed threshold functions.
However, if the two bright outliers are removed from the sample, the
distribution of the remaining 15 FRB is consistent with the Euclidean
hypothesis.
\section{Discussion}
\subsection{ Uncertainties}
This result is subject to the caveat that the uncertainties in the
measured fluxes and widths have not been allowed for.  This is difficult
with the present Catalogue because uncertainties are missing for several of
the bursts, and the meaning of those uncertainties that are in the Catalogue
is unclear.  For example, some uncertainty ranges are very asymmetric about
the nominal values; the implied likelihood distributions must be far from
Gaussian, but are unquantified.  { The fluences and their uncertainty
ranges in the Catalogue are the products of the fluxes, widths and the
limits of flux and width uncertainty ranges (with small deviations for the
lower bound on fluence for a few bursts).  This is questionable because the
flux, width and fluence measurements are not independent; fluence is
constrained independently of the flux and width, but its uncertainty is not
given independently in the Catalogue.  The maximum possible fluence is
overestimated if the maximum flux is multiplied by the maximum width.  Large
uncertainty ranges for a few bursts (particularly FRB 130729, whose width is
given as $15.61\, {+9.98 \atop -6.27}\,$ms) may introduce spurious large
uncertainties in average quantities.  This problem might be addressed by
removing bursts with large uncertainties from the database, but the
decision of which to remove would necessarily be subjective.}

The discrepancy with the Euclidean model is insensitive to uncertainties in
the signal strengths of either the two very bright bursts or the remaining
15 because it is attributable to the absence or deficiency of bursts with
signal strengths intermediate between these two widely separated groups.
For FRB 150807 the quoted uncertainties are small, while for FRB 010724 the
Catalogue contains only lower bounds that we used as actual values; if the
true values were greater than these lower bounds, the discrepancy would be
even larger.
\subsection{ Sensitivities}
\begin{figure}
\centering
\includegraphics[width=0.99\columnwidth]{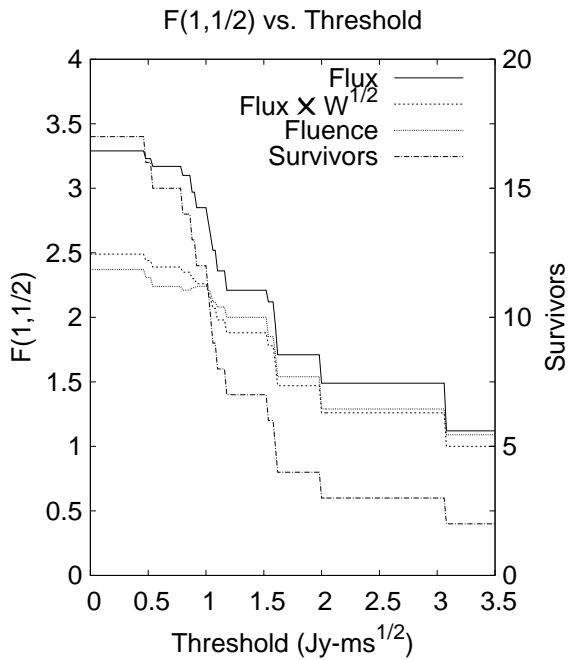}
\caption{\label{thresh}  $F(1,1/2)$ for three measures of FRB signal
strength, as functions of a threshold (of Flux$\times W^{1/2}$).  To
simulate the effect of varying $S_0$, bursts below this threshold are
excluded from the analysis.  The signal strength functions are indicated on
the left ordinate axis; the number of surviving bursts on the right ordinate
axis.  The values of $F(1,1/2)$ do not significantly decrease until half the
bursts are excluded, indicating that the results are not sensitive to
possible variations in $S_0$ in the Catalogue.  $F(1,1/2)$ must decrease as
the weakest bursts are excluded, even if $S_0$ is exactly uniform, so the
decrease does not indicate any variation in $S_0$ among the observations in
the Catalogue.  In contrast to Table~\ref{FRBcat}, here the faintest bursts
are excluded.}
\end{figure}

{ A separate caveat arises from the possibility that {\it different\/}
values of $S_0$ were effectively used in the data analyses, which would
invalidate the derivation in Eq.~\ref{Salpha}.  Even though all Parkes
bursts were observed with the same telescope, data reduction algorithms and
acceptance criteria might have varied.  Data with different $S_0$ cannot be
combined because in a combined dataset the relation between $N$ and $R$
(equivalently, between $N$ and $S$) would then no longer be that implied by
homogeneity and the inverse square law.  Data from the less sensitive
(larger $S_0$) UTMOST and ASKAP cannot be commingled with the Parkes data
because that would introduce a spurious excess of strong bursts.

We can simulate the effect of variable sensitivity by artificially raising
the threshold for acceptance to exclude the weaker bursts in the Catalogue,
as if a stricter criterion (larger $S_0$) were applied to their detection.
This could be done for any signal strength parameter that follows an inverse
square law, but we choose Flux$\times W^{1/2}$ as most closely describing
the detection threshold of observations limited by detector thermal noise.
The 15 weaker Parkes bursts in the Catalogue have values of
Flux$\times W^{1/2}$ ranging from 0.47 to 3.08 Jy-ms$^{1/2}$, with only one
of the 15 above 2 Jy-ms$^{1/2}$.

We therefore repeat the analysis with thresholds ranging from 0 to 3.5
Jy-ms$^{1/2}$ (any threshold below 0.47 Jy-ms$^{1/2}$ admits the full
dataset, while a threshold above 3.08 Jy-ms$^{1/2}$ reduces the dataset to
the two very bright bursts).  The results are shown in Fig.~\ref{thresh}.
The function $F(1,1/2)$ for each of the three measures of FRB signal
strength remains above 2, corresponding to a 98\% significant discrepancy
with the homogeneous Euclidean model, for any threshold below 1.08
Jy-ms$^{1/2}$, at which eight of the 15 weaker bursts are excluded.  The
slowness of the decrease of $F(1,1/2)$ with increasing threshold implies
that the result is insensitive to possible inconsistencies in the acceptance
criteria for bursts ($S_0$).  This result does not require knowledge of the
instrumental sensitivity, unlike the argument of \citet{L07} that the
deficiency of weak bursts was significant because the detection threshold
was far below the signal strength of the (then) one observed burst.}

{ We also simulate the effects of possible burst-to-burst variation
in detection threshold $S_0$ by substituting for the 15 faint bursts a
synthetic set consisting of 15 bursts randomly chosen from the 15 actual
faint bursts.  In the synthetic set one or more of the actual bursts may be
omitted and others represented more than once.  Unlike the thresholded
samples of Fig.~\ref{thresh}, all sets have a total of 17 bursts so there
are no trends resulting from the reduction in number of weak bursts as the
threshold is raised.  Averaging over $10^5$ realizations, we find the
results shown in Table~\ref{FRBcat} as 17 Randomized.  The simulated
uncertainties are not large, and do not affect our conclusions.}
\subsection{Bias}
A final caveat arises from possible bias introduced by the fact that this
study was performed not long after the discovery \citep{R16} of the bright
burst FRB 150807.  If that discovery motivated this study (a question
unanswerable because it depends on human thought processes), the sample was
biased to include a maximal fraction of very bright bursts.  { To estimate
this bias the analysis was repeated excluding FRB 150807, with results
shown in Table~\ref{FRBcat} for 16 bursts.  The weakest constraint is now
obtained from the flux data, and still indicates a 98\% significant
rejection of the homogeneous Euclidean hypothesis.  As the number of
observed bursts increases, any such bias will have less effect.}
\section{Conclusion}
The distribution of FRB in space appears to violate the cosmological
principle that, averaged over sufficiently large scales, the Universe is
homogeneous.  This conflict is resolved if ``sufficiently large'' means on
scales greater than the unknown distances to the two very bright bursts.

If FRB are, roughly, standard candles, and if the repeating FRB 121102 at
$z = 0.193$ is representative of FRB distances in general, then we may
roughly estimate the redshifts of the bright FRB as $\sim 0.03$.  The
assumption that FRB 121102 is representative is unproven; despite Ockham's
Razor to the contrary, it might be a different class of object, as SGR were
distinguished from GRB only many years after their discovery in 1979.  If
FRB 121102 is representative, then the density of FRB sources at $z \lesssim
0.03$ is greater than their mean density at $z \sim 0.2$ by $\sim 50$, the
ratio of the volumes (the 3/2 power of the ratio of their signal strengths)
times the ratio of the numbers (2/15) in the FRB Catalogue.  These redshifts
are small enough that cosmological evolution should have only a minimal
effect on FRB density and the non-Euclidean geometry of space only a minimal
effect on FRB statistics.
\section*{Acknowledgements}
I thank LCK for sharp questions and pointed insights.

\bsp 
\label{lastpage} 
\end{document}